# PNADIS: An automated Peierls-Nabarro Analyzer for DISlocation core structure and slip resistance

S. H. Zhang[1, 2], D. Legut[3] and R. F. Zhang[1, 2, *]

[1]*School of Materials Science and Engineering, Beihang University, Beijing 100191, P. R. China*

[2]*Center for Integrated Computational Engineering (International Research Institute for Multidisciplinary Science) and Key Laboratory of High-Temperature Structural Materials & Coatings Technology (Ministry of Industry and Information Technology), Beihang University, Beijing 100191, P. R. China*

[3]*IT4Innovations Center, VSB-Technical University of Ostrava, CZ-70833 Ostrava, Czech Republic*

\* Corresponding author: zrf@buaa.edu.cn (R. F. Zhang)

**Abstract**

Dislocation is one of the most critical and fundamental crystal defects that dominate the mechanical behavior of crystalline solids, however, a quantitative determination of its character and property in experiments is quite challenging and limited so far. In this paper, a fully automated Peierls-Nabarro (P-N) analyzer named PNADIS is presented; a complete set of the character and property of dislocation can be automatically derived, including the dislocation core structure, Peierls energy and stress, pressure field around dislocation core, solute/dislocation interaction energy, as well as the energy barrier and yield stress at 0K for solid solution strengthening. Furthermore, both one-dimensional (1D) and two-dimensional (2D) P-N models are implemented to meet the demand to analyze the character and property of dislocation for not only simple FCC and HCP structures but also complex crystals. The implementation of this code has been critically validated by a lot of evaluations and tests including 1D P-N model for complex crystals, 2D P-N model for FCC and HCP metals, pressure field around dislocation core, and solid solution strengthening for alloys. We expect that the automated feature of this code would provide a high-efficiency solution for determining the character and property of dislocation.

**Program summary**

*Program title:* PNADIS

*Licensing provisions:* GNU General Public License 3

*Programming language:* MATLAB

*Nature of problem:* To determine automatically the character and property of dislocation, including dislocation core structure, Peierls stress, pressure field around dislocation core and solid solution strengthening, for not only FCC and HCP structures but also complex crystals.

*Solution method:* The generalized stacking fault energy is firstly fitted by Fourier expansion, and meanwhile an appropriate trial function of disregistry vector is chosen. Afterwards, a least square minimization of the difference between elastic resistance and restoring force for one-dimensional Peierls-Nabarro model, or a global minimization of the total dislocation energy via particle swarm optimization or genetic algorithm for two-dimensional Peierls-Nabarro model, will be performed to determine the dislocation core structure of complex crystals, or FCC and HCP structures. Finally, the Peierls stress, pressure field around dislocation core and solid solute strengthening are derived from the calculated dislocation core structure.

# 1. Introduction

Dislocation is one of the most important and fundamental defects that dominate the mechanical properties of crystalline solids [1]. For instance, the stacking fault width of extended dislocations governs the mobility, cross slip, dislocation-dislocation locking, and so on [1]; the maximum lattice resistance to dislocation motion, which is generally defined as Peierls energy or stress [2, 3], is one fundamental quality that describes the crystal plasticity and strength; the solid solute strengthening originates mainly from the "pinning" force of the immobile solute atoms on the dislocation, thus its modelling needs a critical quantification of the solute/dislocation interaction energy [4-6]. Nevertheless, a quantitative determination of the character and property of dislocation in experiments is quite challenging and limited so far. Recently, with the development of modeling methodologies and computational sciences, theoretical investigations on the character and property of dislocations have reignited the great scientific interests due to its enhanced prediction precise; the corresponding methodologies can be divided into two categories: atomistic and continuum descriptions [7].

For the first category, the atomistic description, e.g. flexible boundary conditions [8-10] and dislocation dipole array [11, 12], is that the dislocation cores are characterized explicitly in an atom-by-atom manner [7], in which the atomic structure is determined directly by *ab initio* density functional theory (DFT) calculation or molecular statics/dynamics simulation. Unfortunately, each approach shows some inherent shortcomings: the *ab initio* DFT calculation are very expensive computationally as several hundreds of atoms are required for dislocation simulation albeit they are accurate; while the molecular statics/dynamics simulations are efficient in space scale, but it is often limited by the unavailable reliable empirical potentials.

On the other hand, the continuum description treats a dislocation as continuum object, which makes it possible to consider the dislocation behavior on larger length and longer time scales [7]. For decades, one primary continuum description, *i.e.* the Peierls-Nabarro (P-N) model, has brought considerable interest to study the properties of dislocation due to its simplicity in formulation and efficiency in solution. In the original derivation of P-N model, only the generalized stacking fault energy (GSFE) and elastic modulus are required. Since the proposed elastic-plastic hybrid model [13, 14], the two-dimensional (2D) P-N model has been widely employed to study the dislocation core structures of various metals (e.g. Mg [15, 16], Ni [17], Al [18] and so on) and several transition-metal carbides with B1 structure (e.g. HfC and TaC [19]). And some have been performed via one-dimensional (1D) P-N model for more complex crystals or slip systems, such as BCC structure (e.g. Fe [20]), L1$_2$ Ni$_3$Al [21], perovskite (e.g. SrTiO$_3$ [22] and MgSiO$_3$ [23]) and Mg$_2$SiO$_4$ ringwoodite [24]. More recently, a renewed interest in the P-N model is mostly motivated by the following two facts: i) one more accurate determination of GSFE, owing to the advance of reliable DFT calculation, brings the numerical solution of P-N model into a more realistic level; ii) the limitations of classical continuum P-N (CCPN) model bring the necessary for the newly proposed semidiscrete variational P-N (SVPN) model by Bulatov *et al.* [25], which provides results remarkably similar to those from realistic atomistic simulations.

Though the P-N model has been brought more and more interest and has been employed in various crystals successfully, to best of our knowledge, an automated P-N derivation for the character and property of dislocation has not been implemented in any open-source code available in public so far. Therefore, we here present an automated program named PNADIS: Peierls-Nabarro analyzer for dislocation core structure and slip resistance, in which both 1D and 2D P-N models are supported to meet the demand to analyze the character and property of dislocation for not only simple FCC and HCP structures but also complex crystals. A complete set of the character and properties of dislocation is automatically derived, including the dislocation core structure, Peierls energy and stress, pressure field around dislocation core, solute/dislocation interaction energy, as well as the characteristic bow-out distance, energy barrier and yield stress at 0K for solid solution strengthening. To be noted additionally that an automated procedure is adopted with minimum input parameters in PNADIS code to meet the demands of high-throughput scheme.

In this article, we shall firstly give an overview in Section 2 on the theoretical methods of CCPN and SVPN models to calculate the dislocation core structure, Peierls stress, pressure field around dislocation core and solid solution strengthening. Then, Section 3 presents the workflow and automated scheme of the PNADIS code. Afterwards, several comprehensive evaluations and tests have been performed in Section 4 to validate the implementation and reliability of PNADIS code, including 1D P-N model for complex crystals, 2D P-N model for simple metals, pressure field around dislocation core, and solid solution strengthening for various alloys. In the last Section 5, a brief summary is given with a few remarks on the further development of PNADIS code.

## 2. Overview of theoretical models and methods

### 2.1 Basic concepts in the P-N model

#### 2.1.1 Disregistry vector $u$

The dislocated solid is separated into two elastic parts by the slip plane in the P-N dislocation model. Assigning the displacement fields on either side of the slip plane to be $u^+$ and $u^-$, respectively, the dislocation can be characterized by the disregistry (or misfit) vector $u \equiv u^+ - u^-$. Equivalently, the disregistry distribution across the slip plane is characterized by the disregistry (or misfit) density: $\rho(x) \equiv \nabla u(x)$. Usually, the disregistry vector can be realistically described by the trial function as follows:

$$u(x) = \frac{b}{\pi} \sum_{i=1}^{N} \alpha_i \tan^{-1} \frac{x - d_i}{\omega_i}, \tag{1}$$

or a more accurate trial function by introducing the first-order approximation of $u(x)$ [26]:

$$u(x)=\frac{b}{\pi}\sum_{i=1}^{N}\alpha_i\left[\tan^{-1}\frac{x-d_i}{c_i\omega_i}+(1-c_i)\frac{\omega_i(x-d_i)}{(x-d_i)^2+(c_i\omega_i)^2}\right], \quad (2)$$

where, $\alpha_i$, $d_i$, $\omega_i$ and $c_i$ are variational constants, and $b$ is the Burgers vector. The normalization of $u(x)$ requires that $\sum_{i=1}^{N}\alpha_i=1$ and $\sum_{i=1}^{N}\alpha_i=0$ for the directions along the Burgers vector ($x$ axis) and perpendicular to the Burgers vector within the slip plane ($z$ axis), respectively. These two trial functions correspond to a set of discrete partial dislocations whose Burgers vector, situation and width are $b\alpha_i$, $x_i=d_i$ and $\omega_i$, respectively. Specially, the FCC (111)[1-10] dislocation and HCP (0001)[11-20] dislocation can be well described by only two single partial dislocations as proposed in Ref. [27]:

$$\begin{aligned}u_x(x)&=\frac{b}{2\pi}\left(\tan^{-1}\frac{x-d_x/2}{\omega_x}+\tan^{-1}\frac{x+d_x/2}{\omega_x}\right)+\frac{b}{2}\\ u_z(x)&=\frac{\sqrt{3}b}{6\pi}\left(\tan^{-1}\frac{x-d_z/2}{\omega_z}-\tan^{-1}\frac{x+d_z/2}{\omega_z}\right)\end{aligned}, \quad (3)$$

or by two triplet partial dislocations as suggested in Ref. [18]:

$$\begin{aligned}u_x(x)&=\frac{b}{2\pi}\begin{pmatrix}\alpha_1\tan^{-1}\frac{x-(d_x/2-\Delta d_x)}{\omega_x}+\alpha_2\tan^{-1}\frac{x-d_x/2}{\omega_x}+\alpha_3\tan^{-1}\frac{x-(d_x/2+\Delta d_x)}{\omega_x}\\ +\alpha_4\tan^{-1}\frac{x+(d_x/2-\Delta d_x)}{\omega_x}+\alpha_5\tan^{-1}\frac{x+d_x/2}{\omega_x}+\alpha_6\tan^{-1}\frac{x+(d_x/2+\Delta d_x)}{\omega_x}\end{pmatrix}+\frac{b}{2}\\ u_z(x)&=\frac{\sqrt{3}b}{6\pi}\begin{pmatrix}\beta_1\tan^{-1}\frac{x-(d_z/2-\Delta d_z)}{\omega_z}+\beta_2\tan^{-1}\frac{x-d_z/2}{\omega_z}+\beta_3\tan^{-1}\frac{x-(d_z/2+\Delta d_z)}{\omega_z}\\ +\beta_4\tan^{-1}\frac{x+(d_z/2-\Delta d_z)}{\omega_z}+\beta_5\tan^{-1}\frac{x+d_z/2}{\omega_z}+\beta_6\tan^{-1}\frac{x+(d_z/2+\Delta d_z)}{\omega_z}\end{pmatrix}\end{aligned}, \quad (4)$$

where $u_x(x)$ is the component of the partial dislocations along the $x$ axis, and $u_z(x)$ along $z$ axis. The amplitudes $\alpha_i$ and $\beta_i$ must satisfy the conditions $\sum_{i=1}^{3}\alpha_i=\sum_{j=4}^{6}\alpha_j=1$ and $\sum_{i=1}^{3}\beta_i=-\sum_{j=4}^{6}\beta_j=1$, respectively.

### 2.1.2 Generalized stacking fault energy or γ-surface γ(u)

The GSFE is a critical energetic quantity that depicts the energy variation when one part of crystal is rigidly sliding with respect to the other part along a given crystallographic plane [12], and it can be expressed as:

$$\gamma(u) = \frac{E_{SF}(u) - E_0}{A}, \tag{5}$$

where $E_{SF}$ ($E_0$) is the energy of the slipped (perfect) structure, and $A$ is the area of the slip plane. As demonstrated by Vitek [28], the restoring force introduced in the P-N model is simply the gradient of the GSFE $\gamma(u)$:

$$\tau(u) = -\nabla \gamma(u). \tag{6}$$

The maximum slope $\tau^{\max} = \max\{\tau(u)\}$ namely the ideal slide stress $\tau_{is}$, can be identified as the theoretical shear strength for the rigid interplanar sliding of a crystal along the appropriate slip direction.

The GSFE $\gamma(u)$ can be expressed by Fourier expansion in power of the disregistry vector $u$ in the following equation [15, 27]:

$$\gamma(u) = \sum_G c_G \exp[iGu], \tag{7}$$

where $G = m\frac{2\pi}{a_x}$ and $G = (m\frac{2\pi}{a_x}, n\frac{2\pi}{a_z})$ for 1D and 2D γ-surface, respectively, $m$, $n$=0, ±1, ±2, …, ±∞, and $a_x$ and $a_z$ are the lengths of one period along $x$ and $z$ axes, respectively. Moreover, $\gamma(0)=0$ is required. For convenience, several simplified Fourier series [15, 18] were suggested to express the 2D γ-surface, such as in Ref. [18]:

$$\begin{aligned}
\gamma(u_x,u_z)=c_0&+c_1\left[\cos(2qu_z)+\cos(pu_x+qu_z)+\cos(-pu_x+qu_z)\right]\\
&+c_2\left[\cos(2pu_x)+\cos(pu_x+3qu_z)+\cos(-pu_x+3qu_z)\right]\\
&+c_3\left[\cos(4qu_z)+\cos(2pu_x+2qu_z)+\cos(pu_x-2qu_z)\right]\\
&+c_4\left[\begin{array}{l}\cos(3pu_x+qu_z)+\cos(3pu_x-qu_z)+\cos(2pu_x+4qu_z)\\+\cos(2pu_x-4qu_z)+\cos(pu_x+5qu_z)+\cos(-pu_x+5qu_z)\end{array}\right],\\
&+a_1\left[\sin(pu_x-qu_z)-\sin(pu_x+qu_z)+\sin(2qu_z)\right]\\
&+a_2\left[\sin(2pu_x-2qu_z)-\sin(2pu_x+2qu_z)+\sin(4qu_z)\right]
\end{aligned} \quad (8)$$

where $p=2\pi/a_x$, and $q=2\pi/a_z$.

### 2.1.3 Energy factor K

A simplified form of energy factor K of isotropic solid is defined as

$$K=G\left(\frac{\sin^2\theta}{1-\nu}+\cos^2\theta\right), \quad (9)$$

where $\theta$ corresponds to the angle between the dislocation line and its Burgers vector. Specifically, $\theta$ is equal to 90° and 0° for edge and screw dislocations, respectively. G is the shear modulus and ν is the Poisson's ratio. For anisotropic crystal however, K depends on the slip system and dislocation character, and the anisotropic elastic constants must be accounted for.

### 2.1.4 Interplanar distance Δx

The interplanar distance $\Delta x$ or $a'$ is defined as the shortest distance between two equivalent atomic rows along the dislocation line direction as the absence of a dislocation [2, 25], which is crucial for the SVPN model (see Section 2.3) and the calculation of the Peierls stress (see Section 2.4). Specifically, for the edge and screw dislocations of FCC (111)[1-10] or HCP (0001)[11-20] slip system, $\Delta x$ is equal to $b/2$ and $\sqrt{3}b/2$, respectively.

## 2.2 Classical continuum Peierls-Nabarro model

The total energy of a dislocation is contributed by two parts of elastic energy $E_{elastic}$ and misfit energy $E_{misfit}$, *i.e.* $E_{total}=E_{elastic}+E_{misfit}$. The elastic energy is obtained as a function of the disregistry vector $u(x)$ by integrating the work of the stress in the slip plane [7]:

$$E_{elastic}=-\frac{K}{4\pi}\int_{-\infty}^{\infty}\int_{-\infty}^{\infty}\rho(x)\rho(x')\ln|x-x'|dxdx'. \quad (10)$$

By integrating the GSFE along the displacement direction in the slip plane when introducing the disregistry vector $u(x)$, the misfit energy is expressed as [7]

$$E_{misfit}=\int_{-\infty}^{\infty}\gamma[u(x)]dx. \quad (11)$$

Now the total dislocation energy expression is obtained as follows:

$$E_{total}=-\frac{K}{4\pi}\int_{-\infty}^{\infty}\int_{-\infty}^{\infty}\rho(x)\rho(x')\ln|x-x'|dxdx'+\int_{-\infty}^{\infty}\gamma[u(x)]dx, \quad (12)$$

which is a function of the still unknown disregistry vector $u(x)$.

The unknown $u(x)$ could be determined by solving the equation $\delta E_{total}/\delta u=0$ [29], which corresponds physically to the balance of the elastic resistance $F_{EL}$ with the restoring force $\tau[u(x)]$ and leads to:

$$\frac{K}{2\pi}\int_{-\infty}^{\infty}\rho(x')\frac{1}{x-x'}dx'=\tau[u(x)]. \quad (13)$$

According to the methodology proposed by Joos *et al.* [30], firstly, the previous disregistry trial functions (*i.e.* Eqs. (1-2)) are inserted into the left-hand side of Eq. (13) to give the elastic resistance. Then, the variational constants $\alpha_i$, $d_i$, $\omega_i$ and $c_i$ are determined by a least square minimization of the difference between $F_{EL}$ and $\tau[u(x)]$, which is obtained by deriving the GSFE calculated *ab initio*.

Minimizing the total dislocation energy $E_{total}$ could also determine the unknown $u(x)$. By proposing the trial functions of $u(x)$ as suggested in Eq. (1), the elastic energy is transformed into [29]

$$E_{elastic} = H_{xx}\left[\sum_{i,j}\alpha_{x,i}\alpha_{x,j}\ln\frac{R}{\omega_{x,i}+\omega_{x,j}}-\sum_{i<j}\alpha_{x,i}\alpha_{x,j}\ln\left(1+\frac{r_{ij}^2}{(\omega_{x,i}+\omega_{x,j})^2}\right)\right]\cdot b^2$$

$$+H_{zz}\left[\sum_{i,j}\alpha_{z,i}\alpha_{z,j}\ln\frac{R}{\omega_{z,i}+\omega_{z,j}}-\sum_{i<j}\alpha_{z,i}\alpha_{z,j}\ln\left(1+\frac{r_{ij}^2}{(\omega_{z,i}+\omega_{z,j})^2}\right)\right]\cdot b^2 \quad (14)$$

$$+2H_{xz}\left[\sum_{i,j}\alpha_{x,i}\alpha_{z,j}\ln\frac{R}{\omega_{x,i}+\omega_{z,j}}-\frac{1}{2}\sum_{i,j}\alpha_{x,i}\alpha_{z,j}\ln\left(1+\frac{r_{ij}^2}{(\omega_{x,i}+\omega_{z,j})^2}\right)\right]\cdot b^2$$

Where $\{H_{xx}, H_{zz}\}=\{K_x/4\pi, K_z/4\pi\}$, R is the usual outer cut-off radius and $r_{ij}=|d_j-d_i|$. For the FCC (111)[1-10] dislocation and HCP (0001)[11-20] dislocation, the off-diagonal component $H_{xz}$ vanish, *i.e.* no interaction exists between the edge and screw components [18]. Then, this optimization problem can be solved via particle swarm optimization (PSO) and genetic algorithm (GA).

**2.3 Semidiscrete Variational Peierls-Nabarro model**

Though it has been successfully used in several systems [15, 17, 18], the CCPN model has several limitations [25], such as the discrete nature of the crystalline lattice is not considered, and the elastic energy can be unrealistically high for the narrow dislocation core. Therefore, addressing the limitations of CCPN model, a new SVPN model was presented by Bulatov *et al.* [25]. With replacing the continuum form with the discrete one, the total dislocation energy $E_{total}$ is expressed as

$$E_{total}=\frac{K}{4\pi}\sum_{ij}\chi_{ij}\rho_i\rho_j+\sum_i\gamma[u(x_i)]\cdot\Delta x, \quad (15)$$

$$\chi_{ij}=3/2\varphi_{i,i-1}\varphi_{j,j-1}+\psi_{i-1,j-1}+\psi_{i,j}-\psi_{i-1,j}-\psi_{i,j-1}, \quad (15a)$$

$$\psi_{i,j}=1/2\varphi_{i,j}^2\ln|\varphi_{i,j}|, \quad \varphi_{i,j}=x_i-x_j, \quad (15b)$$

$$\rho_i=(u_i-u_{i-1})/(x_i-x_{i-1}), \quad (15c)$$

where $x_i$ are the reference positions.

## 2.4 Peierls stress

### 2.4.1 Analytical formula

The Peierls stress is defined as the minimum stress for irreversible movement of dislocation with a Burgers vector at 0 K [2, 3]. Assuming a $\tan^{-1}$ profile for the disregistry function $u(x)=(b/\pi)\tan^{-1}(x/\xi)+b/2$ and a sinusoidal form of the restoring force per unit area $\tau(u)=\tau_{is}\sin(2\pi u/b)$, an analytical solution for the Peierls stress $\tau_P$ was given by Joos *et al.* [2]. In the case of wide dislocations, approximately $\xi/a'\gg 1$, the Peierls stress can be obtained as follows

$$\tau_{P,wide}=\frac{Kb}{a'}\exp(-\frac{2\pi\xi}{a'}), \tag{16}$$

where the dislocation half-width $\xi=Kb/(4\pi\tau_{is})$. While for narrow dislocation ($\xi/a'\ll 1$), the Peierls stress can be expressed as follows

$$\tau_{P,narrow}=\frac{3\sqrt{3}}{8}\tau_{is}\frac{a'}{\pi\xi}. \tag{17}$$

### 2.4.2 Discrete dislocation energy approach

When the dislocation locates at the position ε, due to the discrete nature of the crystalline lattice, the misfit energy is rewritten as [2]

$$E_{misfit}(\varepsilon)=\sum_{m}\gamma[u(m\Delta x-\varepsilon)]\cdot\Delta x. \tag{18}$$

The Peierls energy or Peierls barrier $E_P$ represents the average energy change for movement of dislocation from one favorable minimum to a maximum, and is expressed as

$$E_P=\max\{E_{misfit}(\varepsilon)\}-\min\{E_{misfit}(\varepsilon)\}. \tag{19}$$

Accordingly, the Peierls stress can be determined by [2]

$$\tau_P=\max\left\{\frac{1}{b}\frac{dE_{misfit}(\varepsilon)}{d\varepsilon}\right\}. \tag{20}$$

In this method, it is noteworthy that the partial dislocations are implicitly considered to be strongly coupled, therefore, the value of $\tau_P$ calculated via Eq. (20) corresponds to the critical stress required for the whole set of partials overcoming the Peierls barrier without any modification of the core structure [23]. Thus, when one partial climbs the Peierls barrier while the second goes down meantime, it may strongly decrease the calculated Peierls stress, since the total misfit energy in Eq. (18) may be profoundly minimized during such process [23].

**2.5 Pressure field around dislocation core**

The pressure field around dislocation core can be expressed as [31]

$$p(x,y)=-\frac{G(1+v)}{3\pi(1-v)}\sum_{i=1}^{n}\alpha_i b \cdot \frac{y+\text{sgn}(y)\omega_i}{(x-d_i)^2+[y+\text{sgn}(y)\omega_i]^2}, \tag{21}$$

where G and v are the shear modulus and Poisson's ratio, respectively. $n$ is the number of the partial dislocation. The constants $\alpha_i$, $d_i$ and $\omega_i$ are the Burgers vectors, corresponding positions and width of partial dislocation in the disregistry vector formula (*i.e.* Eq. (1)). Via the pressure field, the atoms could be color coded to distinguish the localized stress around the dislocation core.

**2.6 Solid-solution strengthening**

**2.6.1 Solute/dislocation interaction energy**

In PNADIS code, the solute/dislocation interaction energy is determined via 2D P-N model as suggested by Ma *et al.* [31], and is written approximately as the sum of two contributions:

$$E_{\text{int}}(x_i,y_j)=E_{volume}(x_i,y_j)+E_{slip}(x_i,y_j), \tag{22}$$

where $E_{volume}$ and $E_{slip}$ are the interaction energies because of the volumetric misfit of solute atom against the pressure field around dislocation core, and the effect of solute atoms on γ-surface, respectively.

**Volumetric misfit interaction energy:** With the pressure field as Eq. (21), $E_{volume}$ can be calculated by:

$$E_{volume}(x_i, y_j) = -p(x_i, y_j) \times \Delta V, \tag{23}$$

where $\Delta V$ is the extra volume introduced by the solute atom, which could be defined in terms of the lattice parameter [31]:

$$\Delta V = 3V_0^{atom} \cdot \varepsilon_b \text{ and } \varepsilon_b = \left.\frac{d\ln b}{dc}\right|_{c=0}, \tag{24}$$

where $V_0^{atom}$, $b$ and $c$ are the volume of an solvent atom, Burgers vector and solute concentration, respectively. The volumetric misfit parameter $\varepsilon_b$ could be calculated via the method presented in Ref. [32].

**Slip misfit interaction energy:** With calculating the GSFEs of M-X solid solution ($\gamma_{M-X}$) and pure M ($\gamma_M$) on the *j*th slip plane, $E_{slip}$ can be expressed as [31]:

$$E_{slip}(x_i, y_j) = A_{SF} \times \{\gamma_{M-X}[u(x_i, y_j)] - \gamma_M[u(x_i, y_j)]\}, \tag{25}$$

where $u(x_i, y_j)$ is the disregistry at position $x_i$ on the slip plane $y_j$, $A_{SF}$ is defined as the area per solvent atom on the slip plane. With two approximations suggested in Ref. [31], Eq. (25) can be simplified as:

$$E_{slip}(x_i, y_{\pm 1}) = A_{SF} \times \gamma[u(x_i, y_{\pm 1})] \times \varepsilon_s \text{ and } \varepsilon_s = \left.\frac{d\ln\gamma_{I_2}}{dc_{SF}}\right|_{c_{SF}=0}, \tag{26}$$

where $c_{SF}$ is the areal concentration of solute atom within the slip plane, and the slip misfit parameter $\varepsilon_s$ could be calculated via the method presented in Ref. [32].

**2.6.2 Solid solution strengthening model**

The Nabarro-Labusch-Leyson solid solution strengthening model [4-6, 33] is employed in PNADIS code, in which it is assumed that the randomly placed immobile solute atom should bow out the straight dislocation because of the solute/dislocation interaction. The final favorite shape of the bow-out dislocation, which is characterized by the characteristic segment length ($\zeta_c$) and the characteristic bow-out distance ($w_c$), is determined by two competing processed. Firstly, the binding energy ($E_{binding}$) of the dislocation decreases because of favorable solute fluctuations, and can be expressed as a function of the segment length $\zeta$ and bow-out distance $w$ [4-6]:

$$\Delta E_{binding}(\zeta,w)=-\left[\left(\frac{c\zeta}{\sqrt{3}b}\right)^{\frac{1}{2}}\Delta \tilde{E}_p(w)\right]\cdot\left(\frac{L}{2\zeta}\right), \tag{27}$$

where

$$\Delta \tilde{E}_p(w)=\left[2\sum_{ij}(1-\chi(w,y_j))E_{int}(x_i,y_j)\right]^{\frac{1}{2}}. \tag{28}$$

Here, $L$ and $c$ are the length of dislocation and concentration of solute atom, respectively. $E_{int}(x_i,y_j)$ is the solute/dislocation interaction energy with one solute atom locating at $(x_i,y_j)$, and the correlation function

$$\chi(w,y_j)=\frac{\sum_k E_{int}(x_k-w,y_j)E_{int}(x_k,y_j)}{\sum_k E_{int}(x_k,y_j)^2} \tag{29}$$

Secondly, the line energy ($E_{line}$) increases due to the bow-out configuration and can be obtained based on the isotropic line tension model [4-6]:

$$\Delta E_{line}=\Gamma\left(\frac{w^2}{2\zeta}\right)\left(\frac{L}{2\zeta}\right), \tag{30}$$

where $\Gamma$ is the dislocation line tension, which could be determined from the isotropic linear elasticity [34], i.e. $\Gamma_{edge}=\frac{1-2\nu}{2(1-\nu)}\cdot Gb^2$ for edge dislocation, and $\Gamma_{screw}=\frac{1+\nu}{2(1-\nu)}\cdot Gb^2$ for screw dislocation. $\nu$, G and $b$ are the Poisson's ratio, shear modulus and Burgers vector, respectively.

Now, the total energy change $\Delta E_{tot}$ as one straight dislocation of length L goes to the bow-out one of segment length $\zeta$ and bow-out distance $w$ is [4-6]:

$$\Delta E_{tot}(\zeta,w)=\left[\left(\Gamma\frac{w^2}{2\zeta}\right)-\left(\frac{c\zeta}{\sqrt{3}b}\right)^{\frac{1}{2}}\Delta\widetilde{E}_p(w)\right]\cdot\left(\frac{L}{2\zeta}\right). \tag{31}$$

To minimize Eq. (31), the characteristic segment length $\zeta_c$ could be obtained as a function of the bow-out distance $w$:

$$\zeta_c(w)=\left(4\sqrt{3}\frac{\Gamma^2 w^4 b}{c\Delta\widetilde{E}_p^2(w)}\right)^{\frac{1}{3}}, \tag{32}$$

then the total energy change $\Delta E_{tot}$ is expressed as a function of only $w$ [4-6]:

$$\Delta E_{tot}(w,\zeta_c(w))=-\frac{3^{2/3}}{8\cdot 2^{1/3}}\left(\frac{c^2\Delta\widetilde{E}_p^4(w)}{b^2 w^2 \Gamma}\right)^{\frac{1}{3}}L. \tag{33}$$

Minimizing numerically the total energy change per unit length $\Delta E_{tot}/L$, the characteristic bow-out distance $w_c$ is determined, and then the zero-temperature energy barrier $\Delta E_{barrier}$ for moving the dislocation pinned by solute atoms from one favorable configuration to another, can be expressed as [4-6]:

$$\Delta E_{barrier}=\left(\frac{4\sqrt{2}-1}{3}\cdot\frac{3^{5/6}}{2^{5/3}}\right)\left(\frac{cw_c^2\Gamma\Delta\widetilde{E}_p^2(w_c)}{b}\right)^{\frac{1}{3}}, \tag{34}$$

and the corresponding yield stress $\tau_{y0}$ at 0 K is obtained as follows [4-6]:

$$\tau_{y0}=\frac{\pi}{2}\frac{\Delta E_{barrier}}{b\zeta_c(w_c)w_c}=1.01\cdot\left(\frac{c^2\Delta\widetilde{E}_p^4(w_c)}{\Gamma b^5 w_c^5}\right)^{\frac{1}{3}}. \tag{35}$$

## 3. Implementations and workflows

In the section, we describe the workflow and automated scheme of the PNADIS code based on both the CCPN and SVPN models to calculate the character and properties of dislocation, including dislocation core structure, Peierls stress, pressure field around dislocation core and solid solution strengthening listed in Table 1. An automated procedure with minimum input parameters is adapted to meet the demands of high-throughput scheme. The workflow of PNADIS code is schematically shown in Fig. 1 and more details are discussed below:

*Read the input file and load default values*

Before calculating the properties of dislocation via PNADIS code, the input file named infile.m, including the GSFE, elastic moduli and so on, is needed. The PNADIS code will start from reading the infile.m file, then loading its default value if the value of one parameter is not defined in the infile.m file. In the meantime, it is judged whether the value of parameters defined in the infile.m file is wrong or not. If wrong, the PNADIS code will stop with outputting the error information. All the parameters included in PNADIS code are listed in the Table 2, together with a short description and the default value.

*Specify 1D or 2D P-N model*

The type of either 1D or 2D P-N model is chosen for the calculation of dislocation properties, which is determined by whether the GSFE is 1D or 2D. To be noticed that for 2D P-N model, it is mainly used to the FCC (111)[1-10] dislocation and HCP (0001)[11-20] dislocation, e.g. Al [18], Mg [15, 16] and B1-HfC [19], while for 1D P-N model, it is mainly used for the dislocation in more complex crystals or slip systems, such as BCC structure (e.g. Fe [20]), perovskite (e.g. $SrTiO_3$ [22] and $MgSiO_3$ [23]) and $Mg_2SiO_4$ ringwoodite [24]).

*Fit the GSFE data*

In PNADIS code, the GSFE could be fitted via several successful functions, e.g. Eq. (8). Also, the generalized function of Eq. (7) is able to obtain, for which a plane-wave cutoff *k*, *i.e.* $|m|\leq k$ for 1D γ-surface and $\sqrt{m^2+n^2}\leq k$ for 2D γ-surface, is introduced. An example of fitting the γ-surface of Al (111) plane is presented in Fig. 2, together with the difference between the fitting and initial GSFE data.

*Choose the trial function of disregistry vector*

The success of the P-N model attaining a realistic description of dislocation core configuration with a minimum number of parameters is closely dependent on the suitability of trial functions. In the PNADIS code, several trial functions, which have been successfully applied in many dislocation systems [18, 27], can be obtained for 2D P-N model, *i.e.* Eqs. (3-4). Besides these, the generalized trial function of Eq. (1) is also able to obtain for both 1D and 2D P-N model. The more accurate trial function of Eq. (2) with introducing the first-order approximation of *u(x)* can be obtained only for 1D P-N model.

*Optimize*

For 1D P-N model, the restoring force is firstly determined by deriving the GSFE. Then, a least square minimization of the difference between $F_{EL}$ and $\tau[u(x)]$ will be performed to determine the dislocation core structure, which is defined by the variational constants $\alpha_i$, $d_i$, $\omega_i$ and $c_i$ in the trial function of *u(x)*. An example of this process for a SrTiO$_3$ ⟨110⟩{110} edge dislocation is shown in Fig. 3. While for 2D P-N model, the total energy of dislocation based on Eq. (12) for CCPN model and Eq. (15) for SVPN model is firstly determined, then by minimizing the total energy $E_{total}$ of dislocation via PSO or GA, the dislocation core structure is determined. It must be noted that the lower and upper bounds, and the initial values of these variational constants are very crucial for whether the global minimum could be found. In PNADIS code, the lower and upper bounds, and the initial values could be set in the input file of infile.m.

*Calculate the Peierls stress and pressure field*

After that, the Peierls stress, pressure field around dislocation core and solid solute strengthening can be determined based on the results of dislocation core structure. In PNADIS code, the Peierls energy and stress could be calculated via two methods as introduced in Section 2.4. Examples of determining the Peierls stress of Pd via the analytical formula and the discrete dislocation energy approach based the results of 1D P-N model are presented in Fig. 4. The GSFE and energy factor in Ref. [35] were used. The pressure field, which is as the input parameter of solid solution strengthening model, can be also calculated by our code using Eq. (21). In PNADIS code, it is also able to continue calculating the Peierls stress, pressure field and solid solution strengthening based on the previous results of dislocation core structure by loading the pnadis.mat file, instead of calculating the dislocation core structure once again.

*Calculate the solid solution strengthening*

The solute/dislocation interaction energy is firstly determined on the basis of the results of dislocation core structure and pressure field. Then, it is used to determine the two characteristic parameters of $\zeta_c$ and $w_c$ for a bow-out dislocation in a randomly distributed solid solution. In turn, $\zeta_c$ and $w_c$ are employed to determine the energy barrier ($\Delta E_{barrier}$) and the corresponding yield stress ($\tau_{y0}$) at 0K. After ending of all calculating progresses, a data file named pnadis.mat will be output, including the values of all input and output parameters.

## 4. Evaluations and discussions

### 4.1 1D P-N model for complex crystals

Table 3 presents the calculated partial dislocation separation distance ($D_{SF}$), Peierls energy ($E_P$) and Peierls stress ($\tau_P$) of Si, Pd, Al, SrTiO$_3$, Mg$_2$SiO$_4$ and MgSiO$_3$, together with the previous theoretical values [22-24, 30, 35]. The trial function of Eq. (1) was employed, and the discrete dislocation energy approach was used to determine the Peierls stress. Note that the values of energy factor and GSFE in the literature were used to compare the results calculated via PNADIS code with the previous results in the literature more realistically. It is shown that all calculated values by PNADIS code are in reasonable agreement with the previous theoretical values [22-24, 30, 35], confirming the validity of PNADIS code for 1D P-N model. An example of determining the dislocation core structure of the SrTiO$_3$ ⟨110⟩{110} edge dislocation via 1D P-N model is presented in Fig. 3. It is found that the plateau on the GSFE curve results in a significant core spreading with $D_{SF}$=11.8 Å, and an agreement is reached with the previous theoretical value of 13.2 Å (as shown in Table 3). However, the two particle dislocations are not individualized. As shown in Fig. 3b, the fitting values of $F_{EL}$ agree completely with the calculated restoring force, which indicates the accuracy of the PNADIS code. Fig. 4a illustrates a schematic of the discrete dislocation energy approach to determine the Peierls stress of Pd via 1D P-N model. The GSFE and energy factor in Ref. [35] were used. It is indicated that the Peierls barrier $E_P$ and Peierls stress $\tau_P$ are determined to be $1.62 \times 10^{-12}$ J/m and 136 MPa, which are very close to the reported values of $1.76 \times 10^{-12}$ J/m and 147 MPa in Ref. [35], respectively (as listed in Table 3).

In addition, the analytical formula method is also applied to calculate the Peierls stress of Pd and Al and the results are shown in Table 3. Our results show a good agreement with previous theoretical values [35] too. The specific process of determining the Peierls stress of Pd via the analytical formula is presented in Fig. 4b, with using the GSFE and energy factor in Ref. [35]. It is seen that the dislocation half-width ξ is determined to be 1.48 Å with an ideal slide stress $\tau_{is}$ of 11.4 GPa. Ultimately, the Peierls stress is determined to be 174 MPa, which is very close to the reported value of 173 MPa in Ref. [35], as shown in Table 3.

### 4.2 2D P-N model for FCC and HCP metals

Tables 4 and 5 list the calculated GSFEs, geometrical parameters of dislocation cores and Peierls stress of all FCC and HCP metals via both of CCPN and SVPN models, together with the experimental data [35-54] and other theoretical values [8, 9, 15, 17, 52, 54-89]. The trial function of Eq. (3) was employed, and the discrete dislocation energy approach was used to determine the Peierls stress. As $d_z$ and $w_z$ are nearly equal to $d_x$ and $w_x$, respectively, only $d_x$ and $w_x$ are listed. Note that the energy factors and GSFEs of FCC and HCP metals employed in this paper are calculated *ab initio* via the method in Ref. [16] and the elastic properties were determined via the AELAS code [90]. It is found that our calculated GSFEs show a good agreement with the experimental data [35-48] and other theoretical values [55-67]. An example of fitting the γ-surface of Al (111) plane is presented in Fig. 2, together with the difference between the fitting and initial GSFE data. It is seen that the PNADIS code could fit the GSFE data calculated *ab initio* well.

The predicted geometrical parameters of dislocation cores and Peierls stress listed in Tables 4 and 5 indicate that the full FCC (111)[1-10] and HCP (0001)[11-20] dislocations dissociate into two partial dislocations with a planar stacking fault in between for all of FCC and HCP metals. The predicted values via our PNADIS code lie well within the broad ranges of those reported previously in the literature, including the experimental data [49-54] and other theoretical values [8, 9, 15, 17, 52, 54, 68-89]. And the dislocation core structures and Peierls stresses of Sc, Y, Hf, Re and Os are reported for the first time, for which no experimental or theoretical data has been reported to the best of our knowledge. Furthermore, it is found that the calculated $D_{SF}$ (*i.e.* $d_x$) via the CCPN and SVPN models are nearly equal, while the calculated $w_x$ via the SVPN model is in general smaller than that via the CCPN model, which results in the much larger Peierls stresses calculated via the SVPN model in comparison with those calculated via the CCPN model.

**4.3 Pressure field around dislocation core**

Fig. 5 presents the pressure field (in GPa) around dislocation core for both the components along *x* (*i.e.* [1-10]) and *z* (*i.e.* [11-2]) axes of Al (111)[1-10] edge and screw dislocations calculated by PNADIS code. The differences between the components of partial dislocation along *x* and *z* axes can be easily found from Fig. 6. The atoms are color coded as a function of the localized pressure value, and it is clearly found that two partial dislocations (denoted by "⊥") are separated by a planar stacking fault, corresponding to a full edge or screw dislocation dissociates into two Shockley partials on the (111) plane of FCC structures or (0001) plane of HCP structures (as shown in Fig. 6). The different view directions of the pressure field in Fig. 5 for the edge and screw dislocations are shown in Fig. 6 as the red hollow arrows.

**4.4 Solid solution strengthening for Al alloys**

Table 6 lists the characteristic bow-out distance ($w_c$), and the predicted $\Delta E_{barrier}/c^{1/3}$ and $\tau_{y0}/c^{2/3}$ at 0K of Al alloys with various solute elements calculated via PNADIS code. The line tensions, and the volumetric and slip misfit parameters in the Ref. [31] were used. The agreement between our calculated results and the previous theoretical values [31] provides a validation of the implementation of PNADIS code for the solid solution strengthening. Fig. 7 shows the energy barrier ($\Delta E_{barrier}/c^{1/3}$) and corresponding yield stress ($\tau_{y0}/c^{2/3}$) at 0K of the edge and screw dislocations as a function of the misfit parameters $\varepsilon_b$ and $\varepsilon_s$. It is found that the energy barrier $\Delta E_{barrier}$ of edge dislocation is much larger than that of screw dislocation, while the yield stress $\tau_{y0}$ at 0K of edge dislocation is lower than that of screw dislocation. This is mainly due to the much shorter characteristic bow-out distance of screw dislocation comparing with that of edge dislocation [31] (as shown in Table 6). Furthermore, both $\Delta E_{barrier}$ and $\tau_{y0}$ at 0K strongly depend on $\varepsilon_b$ for both the edge and screw dislocations but have little relation with $\varepsilon_s$, which is also found on the solid solution strengthening of Mg alloy [32].

The correlation function $\chi(w,y_i)$ as a function of $w$ and $y_i$ for an edge dislocation in the Al-Mg alloy, is performed and shown in Fig. 8a. It is indicated that for any finite $w$, $\chi(w,y_i)$ increases with the increasing $|y_j|$ and as $|y_j|\to+\infty$, the $\chi(w,y_j)\to 1$. An agreement is obtained between the results calculated via our PNADIS code and the previous results of Al-Cr [4] and Al-Mg [6]. Fig. 8b also presents the normalized energy change per unit length $\Delta E_{tot}/Lc^{2/3}$ as a function of the bow-out distance $w$ for an edge dislocation in the Al-Mg alloy. It is seen that the total energy change decreases sharply and then increases gradually with the increasing bow-out distance $w$. The characteristic bow-out distance $w_c$ is determined as 22.1 Å. It is found that there is a similar shape for the curve of $\Delta E_{tot}/Lc^{2/3}$ vs. $w$ with the previous results of those for Al-Cr [4] and Al-Mg [6]. All of those indicate the successful implementation of PNADIS for the solid solution strengthening.

## 5. Conclusions

In summary, the implementation and validation of PNADIS code, an automated Peierls-Nabarro analyzer for dislocation core structure and slip resistance including both 1D and 2D P-N models, have been presented. The automated feature demonstrates its promise as an effective tool with high efficiency for deriving the character and property of dislocation, including dislocation core structure, Peierls energy and stress, pressure field around dislocation core, and solid solution strengthening. We are currently advancing the PNADIS code to implement the modified SVPN model by considering the gradient energy [91] and nonlocal energy [92, 93], and the more accurate method to determine the Peierls stress by adding a term describing the interaction of the applied stress on the total energy of dislocation [94].

**Acknowledgements**

This work is supported by the National Natural Science Foundation of China (NFSC) with No.

51672015, National Key Research and Development Program of China (Nos. 2016YFC1102500 and 2017YFB0702100), "111 Project" (No. B17002), National Thousand Young Talents Program of China, and Fundamental Research Funds for the Central Universities. D.L. was supported by project IT4Innovations-path to exascale No.CZ.02.1.01/0.0/0.0/16_013/0001791 and grants No. 17-27790S of Czech Science Foundation and No. LQ1602 of National Programme of Sustainability.

**Tables**

Table 1. Properties derived from the P-N model in PNADIS code.

| Property | Unit | Description | Equation |
| --- | --- | --- | --- |
| $d_i$ | Å | The position of partial dislocation | $d_i$ in Eqs. (1-2) |
| $\omega_i$ | Å | The width of partial dislocation | $\omega_i$ in Eqs. (1-2) |
| $\tau_{is}$ | GPa | The ideal slide stress | $\tau_{is}=\max\{\tau(u)\}$ |
| $\xi$ | Å | The half width of dislocation | $\xi=Kb/(4\pi\tau_{is})$ |
| $\tau_{P, wide}$ | MPa | The Peierls stress calculated via the analytical formula for $\xi/a' \gg 1$ | Eq. (16) |
| $\tau_{P, narrow}$ | MPa | The Peierls stress calculated via the analytical formula for $\xi/a' \ll 1$ | Eq. (17) |
| $E_P$ | $10^{-10}$J/m | The Peierls energy calculated via discrete dislocation energy approach | Eq. (19) |
| $\tau_P$ | MPa | The Peierls stress calculated via discrete dislocation energy approach | Eq. (20) |
| $p(x,y)$ | GPa | The position-dependent pressure filed around dislocation core | Eq. (21) |
| $E_{int}(x,y)$ | eV | The position-dependent solute/dislocation interaction energy | Eq. (22) |
| $w_c$ | Å | The characteristic bow-out distance in the solid solution strengthening model | See main text |
| $\Delta E_{barrier}$ | eV | The energy barrier hindering the dislocation motion at 0 K in the solid solution strengthening model | Eq. (34) |
| $\tau_{y0}$ | MPa | The corresponding yield stress at 0 K in the solid solution strengthening model | Eq. (35) |

Table 2. Overview of all input parameters currently supported in PNADIS code, together with a short description and the default value.

| Name | Description | Default Value |
| --- | --- | --- |
| system | System name | - |
| filepath | The absolute path to save the result files | pwd |
| misdim | The dimension of P-N model: 1D or 2D | - |
| Dislocation_Core_Structure | Calculating the dislocation core structure or NOT | TRUE |
| Ux, Uz, SFE | The data of GSFE and normalized disregistry vector | - |
| Nmis | Which trial function of disregistry vector to employ | 1D P-N model: 3<br>2D P-N model: 0 |
| mis_1st | Considering the first-order approximation in the trial function of disregistry vector or NOT | FALSE |
| BurVect | The value of Burgers vector | - |
| shear_modulus, poisson_ratio | The values of shear modulus and Poisson's ratio | - |
| mistype | The dislocation type: edge or screw | e |
| fitcut | Which GSFE fitting function to employ | 0 |
| pnmode | Mode of P-N models: CCPN or SVPN | 1 |
| Inpas | Interplanar distance $\Delta x$ | - |
| dax | The reference position $x_m$ is defined as $x_m = m\Delta x + dax$ for SVPN model ($m = 0, \pm 1, \pm 2, \cdots, \pm \infty$) | - |
| Xcoef_range, Xdist_range, Xwid_range, Xalpha_range, dx_range | Matrix of the lower and upper bounds, and initial values of each unknown variable | - |
| minimethod | The minimization method for 2D P-N model: PSO or GA | 2d_PSO |
| fitmethod | The mothed for fitting the restoring force | 1d_lsq_curvefit |
| PopulationSize, MaxIterations | Size of the population and maximum number of iteration for PSO and GA | 1000 |
| Peierls_Stress | Calculating the Peierls stress or NOT | FALSE |
| pnstrmethod | The method to calculate the Peierls stress | 2 |
| Pressure_Field | Calculating the pressure field or NOT | FALSE |
| pressfld_latxz, pressfld_latyy | The lattice matrix for pressure field | - |
| dis_component_xz | Which component of the dislocation to calculate: x or z | x |
| Solution_Strengthening | Calculating the solid solution strengthening or NOT | FALSE |
| Einteraction_Plot | Plotting the position-dependent solute/dislocation interaction energy or NOT | FALSE |
| eb_input, es_input | The input volumetric misfit and slip misfit parameters | - |
| LatC_spacing | The interlayer spacing along FCC [111] or HCP [0001] direction | - |
| coefV | The coefficient to calculate the extra volume | - |
| solute_concentration | Concentration of solute atom | 1.0000 |

Table 3. Results of the 1D P-N model for edge and screw dislocations of various materials (*i.e.* Si, Pd, Al, SrTiO$_3$, Mg$_2$SiO$_4$ and MgSiO$_3$), together with the previous theoretical values. The values of energy factor (in GPa) and GSFE in the literatures were employed. D$_{SF}$ (in Å) is the partial dislocation separation distance, E$_P$ (in 10$^{-12}$J/m) corresponds to the Peierls energy, and τ$_P$ (in MPa) is the Peierls stress.

| Str. | Slip system | K$_{edge}$ | D$_{SF}$ | E$_P$ | τ$_{P,e}$ | K$_{screw}$ | D$_{SF}$ | E$_P$ | τ$_{P,s}$ | Note |
|---|---|---|---|---|---|---|---|---|---|---|
| Si | Glide | | | | | | | 6.00×10$^3$ | 1.27×10$^6$ | This work |
| | | | | | | 64 | | 6.00×10$^3$ | 1.45×10$^6$ | Ref. [30] |
| | Shuffle | | | | | | | 234 | 6.6×10$^3$ | This work |
| | | | | | | 64 | | 240 | 7.6×10$^3$ | Ref. [30] |
| Pd | [110] unrel | | | 1.62 | 136, 174$^a$ | | | | | This work |
| | | 77.78 | | 1.76 | 147, 173$^a$ | | | | | Ref. [35] |
| Al | [110] unrel | | | 4.56 | 353, 264$^a$ | | | | | This work |
| | | 39.89 | | 4.16 | 330, 317$^a$ | | | | | Ref. [35] |
| SrTiO$_3$ | <100>{011} | | | 28.2 | 582 | | | 648 | 9.7×10$^3$ | This work |
| | | 143.09 | | 28.8 | 600 | 110 | | 657 | 9.9×10$^3$ | Ref. [22] |
| | <110>{110} | | 11.8 | 0.35 | 3.5 | | 8.9 | 0.36 | 5.1 | This work |
| | | 143.09 | 13.2 | 0.4 | 4 | 109.25 | 9.9 | 0.4 | 6 | Ref. [22] |
| | <110>{001} | | | 133 | 1.3×10$^3$ | | | 88.4 | 1.17×10$^3$ | This work |
| | | 143.09 | | 115 | 1.2×10$^3$ | 109.25 | | 77.6 | 900 | Ref. [22] |
| Mg$_2$SiO$_4$ | 1/2<110>{001} | | 14.2 | 417 | 5.2×10$^3$ | | 10.5 | 711 | 7.3×10$^3$ | This work |
| | | 158 | 13 | 512 | 6.0×10$^3$ | 117 | 10 | 464 | 6.0×10$^3$ | Ref. [24] |
| | 1/2<110>{111} | | 28 | 231 | 2.2×10$^3$ | | 21.3 | 311 | 3.6×10$^3$ | This work |
| | | 154 | 26 | 144 | 1.5×10$^3$ | 117 | 20 | 336 | 4.0×10$^3$ | Ref. [24] |
| MgSiO$_3$ | (010)[001] | | | | | | 10.4 | 812 | 1.75×10$^4$ | This work |
| | | | | | | 278.6 | 9.85 | 864 | 1.85×10$^4$ | Ref. [23] |

$^{a)}$ The results calculated via the analytical formula.

Table 4. First-principles predicted stable and unstable stacking fault energies $\gamma_{I2}$ and $\gamma_{UI2}$ (in mJ/m$^2$), geometrical parameters of dislocation cores ($d_x$ and $w_x$ in Å) and Peierls stresses $\tau_P$ (in MPa) calculated via both CCPN and SVPN models for FCC metals, together with the previous experimental and theoretical values. A trial function of Eq. (3) was employed and $b$ is the Burgers vector.

| FCC | $\gamma_{I2}$ | $\gamma_{UI2}$ | $D_{SF}$ | $w_x$ | Literature values of $d$ | $\tau_P$ (MPa) |
|---|---|---|---|---|---|---|
| Ag | 18.82 | 98.63 | 12.19b[a], 12.75b[b] | 0.55b[a], 0.80b[b] | 13.1b[c] [68], 11.0b[d] [69] | 6.37[a], 1.48[b] |
|  | 17 [55], 18±3[e] [36], 16[e] [35] | 111 [55] |  |  |  | <9[e] [49], 0.60[e] [50] |
| Al | 123.12 | 162.05 |  |  |  | 4.16[a], 0.005[b] |
|  | 146 [56], 158 [57], 126 [58], 167±33[e] [36], 120[e] [37], 166[e] [38] | 178 [56], 175 [57], 169 [58] | 3.55b[a], 3.54b[b] | 0.62b[a], 0.89b[b] | 6.8[c] [70], 9[f] [71], 9[f] [72], 3.2[f] [73], 6.2[e] [51], 8.5[f] [74], 3.5[f] [75], 3.6b[c] [76], 8.5[f] [52], 8.0[e] [52], 9.5[g] [8], 2.0b[d] [69] | <1.4[e] [49], 1.41[e] [50], 1.8[e] [53], 2.6[f] [71], 3.5[f] [74], 35[c] [75] |
| Au | 27.72 | 77.70 | 8.03b[a], 8.11b[b] | 0.58b[a], 0.84b[b] | 39[c] [54], 32.5[e] [54], 10.2b[c] [68], 9.0b[d] [69] | 0.62[a], 0.73[b] |
|  | 27 [55], 37±8[e] [36] | 94 [55] |  |  |  | <0.9[e] [49] |
| Cu | 38.66 | 164.22 |  |  |  | 1.81[a], 0.78[b] |
|  | 43 [55], 41 [59], 38 [56], 39 [57], 37 [58], 61±17[e] [36], 45[e] [39], 42[e] [40], 40[e] [41], 50[e] [42], 41[e] [43] | 175 [55], 164 [56], 158 [57], 180 [58] | 12.15b[a], 12.75b[b] | 0.60b[a], 0.88b[b] | 36[c] [70], 52.5[f] [72], 40[f] [77], 5.5b[c] [76], 10.5b[f] [52], 9.0b[d] [69] | <0.28[e] [49], 0.50[e] [50] |
| Ir | 362.20 | 653.69 | 5.50b[a], 5.69b[b] | 0.67b[a], 1.00b[b] | 5.6b[c] [76], 16[c] [54], 12.5[e] [54], 7.0b[d] [69] | 11.43[a], 0.68[b] |
|  | 359 [55], 390±90[e] [36] | 753 [55] |  |  |  |  |
| Ni | 137.26 | 282.36 | 7.05b[a], 7.25b[b] | 0.66b[a], 0.98b[b] | 3.0b-5.2b[f] [78], 14.9[f] [79], 8.9b[c] [68], 25[f] [80], 5.64b[c] [17], 8.0b[d] [69] | 5.54[a], 0.39[b] |
|  | 137 [56], 127 [60], 125[e] [35] | 278 [56], 263 [60] |  |  |  | 6.40[e] [50] 4.9[c] [79] |
| Pd | 142.65 | 215.68 | 3.27b[a], 3.04b[b] | 0.45b[a], 0.61b[b] | 5.4b[c] [76], 3.8b[d] [81], 17.6[c] [82], 4.6b[c] [83], 5.0b[c] [69] | 59.61[a], 6.85[b] |
|  | 122 [55], 177±3[e] [36], 180[e] [35] | 215 [55] |  |  |  | 0.98[f] [82], 2.6[c] [82] |
| Pt | 293.10 | 301.71 | 2.81b[a], 2.74b[b] | 0.59b[a], 0.84b[b] | 5.5b[c] [76], 4.0b[d] [69] | 11.11[a], 1.18[b] |
|  | 282 [55] | 311 [55] |  |  |  |  |
| Rh | 201.43 | 457.53 | 6.95b[a], 7.14b[b] | 0.66b[a], 0.99b[b] | 9.0b[d] [69] | 15.74[a], 0.33[b] |
|  | 205 [55] | 489 [55] |  |  |  |  |
| Pb | 38.64 | 53.26 | 2.53b[a], 2.47b[b] | 0.42b[a], 0.57b[b] |  | 18.77[a], 3.94[b] |
|  | 25[e] [36] |  |  |  |  | 0.22[e] [53] |

[a)] 2D SVPN model in this work;

[b)] 2D CCPN model in this work;

[c)] P-N model in previous literature;

[d] Phase field dislocation dynamics;

[e] Experiment;

[f] Molecular statics/dynamics simulation;

[g] Ab initio DFT calculation.

Table 5. First-principles predicted stable and unstable stacking fault energies $\gamma_{I2}$ and $\gamma_{UI2}$ (in mJ/m$^2$), geometrical parameters of dislocation cores ($d_x$ and $w_x$ in Å) and Peierls stresses $\tau_P$ (in MPa) calculated via both CCPN and SVPN models for HCP metals, together with the previous experimental and theoretical values. A trial function of Eq. (3) was employed and $b$ is the Burgers vector.

| HCP | $\gamma_{I2}$ | $\gamma_{UI2}$ | $D_{SF}$ | $w_x$ | Literature values of $d$ | $\tau_P$ |
|---|---|---|---|---|---|---|
| Ti | 309.34 | 387.66 | | | | 48.17$^a$, 8.88$^b$ |
| | 287 [61], 306 [62], 310 [63], 292 [64], 319 [65], 300$^d$ [44] | 394 [61], 401 [62] | 1.65b$^a$, 1.45b$^b$ | 0.52b$^a$, 0.72b$^b$ | 5b$^c$ [84], 4b$^c$ [85] | |
| Mg | 33.85 | 93.57 | | | | 11.06$^a$, 0.15$^b$ |
| | 21 [61], 33.8 [66], 36 [67], 78±15$^d$ [45], <50$^d$ [46], >90$^d$ [47], 60$^d$ [48] | 88 [61], 87.6 [66], 92 [67] | 7.34b$^a$, 7.65b$^b$ | 0.54b$^a$, 0.78b$^b$ | 16.7$^e$ [9], 12.7$^c$ [86], 14.3$^c$ [87], 11.5b$^c$ [85], 7.00b$^f$ [15], 3.06b$^c$ [15] | <0.9$^d$ [49], 8.1$^d$ [50], 0.3$^c$ [9], 14$^c$ [9] |
| Be | 520.86 | 847.5 | 1.23b$^a$, 1.79b$^b$ | 0.33b$^a$, 0.45b$^b$ | 3b$^c$ [85] | 566.34$^a$, 143.00$^b$ |
| | 557 [61], 560 [63], 581.46 [48] | 937 [61] | | | | 95.9$^d$ [50] |
| Zr | 221.99 | 258.56 | 2.14b$^a$, 1.89b$^b$ | 0.63b$^a$, 0.93b$^b$ | 4b$^c$ [88] | 8.60$^a$, 0.28$^b$ |
| | 223 [61], 230 [63], 228 [65] | 261 [61] | | | | 1.6$^c$ [89], 1.2$^c$ [88] |
| Zn | 109.43 | 113.93 | 3.80b$^a$, 3.74b$^b$ | 0.73b$^a$, 1.11b$^b$ | | 0.48$^a$, 0.01$^b$ |
| | 102 [61], 100 [63] | 120 [61] | | | | <1.0$^d$ [49], 0.65$^d$ [50] |
| Sc | 176.49 | 243.49 | 3.52b$^a$, 2.51b$^b$ | 0.44b$^a$, 0.61b$^b$ | | 84.55$^a$, 14.37$^b$ |
| | 180 [63] | | | | | |
| Y | 84.83 | 180.06 | 4.38b$^a$, 4.39b$^b$ | 0.45b$^a$, 0.63b$^b$ | | 54.10$^a$, 11.23$^b$ |
| | 100 [63] | | | | | |
| Hf | 325.20 | 431.63 | 2.14b$^a$, 2.02b$^b$ | 0.50b$^a$, 0.70b$^b$ | | 48.80$^a$, 7.44$^b$ |
| | 330 [63] | | | | | |
| Re | 201.98 | 656.09 | 8.41b$^a$, 8.54b$^b$ | 0.64b$^a$, 0.98b$^b$ | | 44.03$^a$, 2.14$^b$ |
| | 140 [63] | | | | | |
| Os | 744.83 | 1266.16 | 3.56b$^a$, 3.68b$^b$ | 0.49b$^a$, 0.68b$^b$ | | 270.88$^a$, 4.36$^b$ |
| | 890 [63] | | | | | |

$^{a)}$ 2D SVPN model in this work;

$^{b)}$ 2D CCPN model in this work;

$^{c)}$ Molecular statics/dynamics simulation;

$^{d)}$ Experiment;

$^{e)}$ Ab initio DFT calculation;

$^{f)}$ P-N model in previous literature.

Table 6. The characteristic bow-out distance $w_c$ (in Å), the predicted $\Delta E_{barrier}/c^{1/3}$ (in eV) and $\tau_{y0}/c^{2/3}$ (in MPa) at 0K of Al alloys with various solute elements calculated via PNADIS code, using the line tensions, and the volumetric ($\varepsilon_b$) and slip ($\varepsilon_s$) misfit parameters in the Ref. [31].

| | $\varepsilon_b$ | $\varepsilon_s$ | Edge | | | Screw | | | Note |
|---|---|---|---|---|---|---|---|---|---|
| | | | $w_c$ | $\Delta E_{barrier}/c^{1/3}$ | $\tau_{y0}/c^{2/3}$ | $w_c$ | $\Delta E_{barrier}/c^{1/3}$ | $\tau_{y0}/c^{2/3}$ | |
| Al-Mg | | | 22.1 | 5.79 | 275 | 5.33 | 1.52 | 364 | This work |
| | 0.104 | -0.38 | 17.1 | 3.96 | 274 | 4.95 | 1.44 | 411 | Ref. [31] |
| Al-Li | | | 15.1 | 1.68 | 72 | 8.0 | 1.37 | 88 | This work |
| | -0.007 | -0.68 | 11.4 | 1.47 | 128 | 9.9 | 0.97 | 185 | Ref. [31] |
| Al-Cu | | | 22.2 | 5.92 | 282 | 5.22 | 1.49 | 373 | This work |
| | -0.107 | 0.22 | 17.1 | 4.01 | 280 | 4.95 | 1.44 | 411 | Ref. [31] |
| Al-Si | | | 20.0 | 3.36 | 124 | 6.06 | 1.17 | 148 | This work |
| | -0.049 | -0.53 | 14.3 | 2.25 | 152 | 4.95 | 1.06 | 222 | Ref. [31] |
| Al-Cr | | | 22.2 | 9.01 | 655 | 5.24 | 2.29 | 867 | This work |
| | -0.201 | 0.49 | 17.1 | 6.10 | 648 | 4.95 | 2.20 | 957 | Ref. [31] |
| Al-Mn | | | 22.1 | 10.18 | 847 | 5.31 | 2.65 | 1120 | This work |
| | -0.242 | 0.84 | 17.1 | 6.94 | 840 | 4.95 | 2.52 | 1257 | Ref. [31] |
| Al-Fe | | | 22.0 | 10.42 | 892 | 5.34 | 2.74 | 1179 | This work |
| | -0.251 | 0.96 | 17.1 | 7.14 | 888 | 4.95 | 2.60 | 1338 | Ref. [31] |

**Figures**

Fig. 1. Workflow of the PNADIS code.

Fig. 2. Example of (a) fitting the γ-surface of Al (111) plane using Eq. (8), together with (b) the difference between the fitting and initial GSFE data.

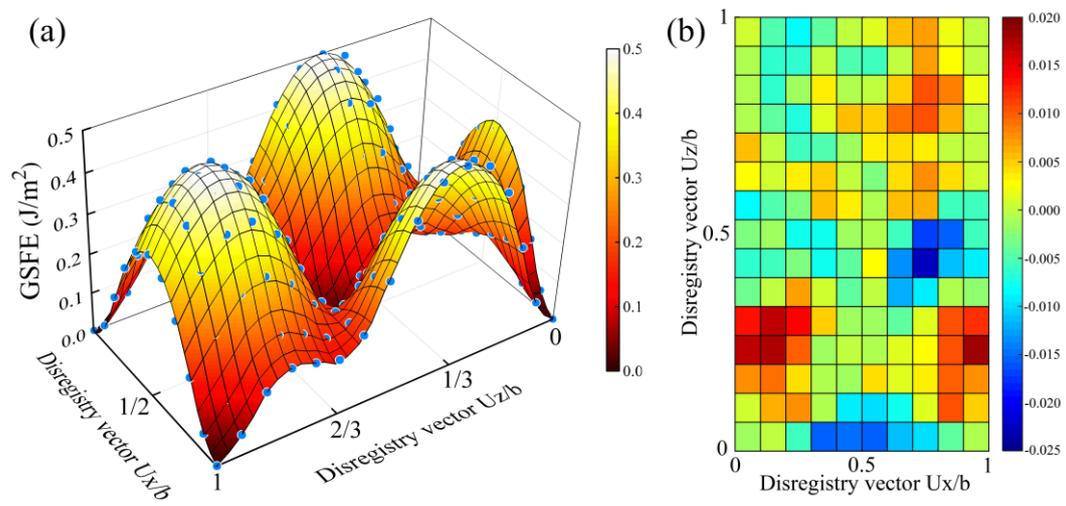

Fig. 3. Example of determining the dislocation core structure of the SrTiO$_3$ ⟨110⟩{110} edge dislocation via 1D P-N model. (a) GSFEs and the derived restoring force vs. *u/b*. (b) A least square minimization of the difference between $F_{EL}$ (solid curve) and $\tau[u(x)]$ (open circles). (c) Disregistry *u/b* and misfit density ρ as a function of *x/b*.

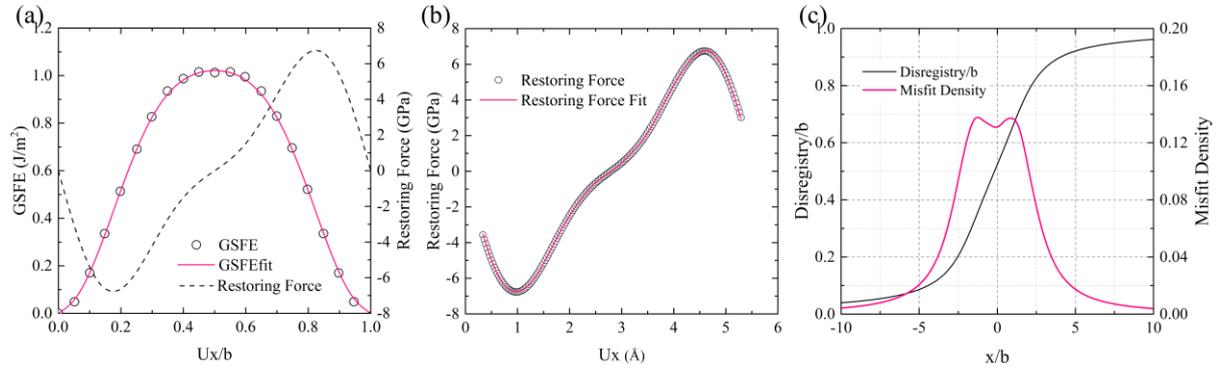

Fig. 4. (a) A schematic of the discrete dislocation energy approach to determine the Peierls stress of Pd via 1D P-N model, using the GSFE and energy factor in Ref. [35]. The Peierls energy $E_P$ and Peierls stress $\tau_P$ are determined to be $1.62 \times 10^{-12}$ J/m and 136 MPa, respectively. (b) A schematic of determining the Peierls stress of Pd via the analytical formula. The GSFE and energy factor in Ref. [35] were used. The Peierls stress is determined to be 174 MPa.

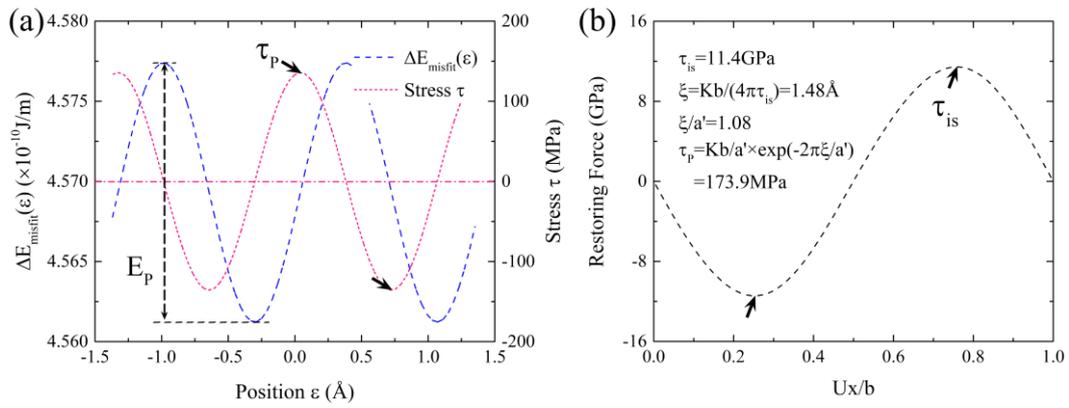

Fig. 5. The pressure field (in GPa) around dislocation core for both the components along *x* (*i.e.* [1-10]) and *z* (*i.e.* [11-2]) axes of Al (111)[1-10] edge and screw dislocations calculated by PNADIS code. The atoms are color coded as a function of the localized pressure value, and the dislocations are denoted by "⊥".

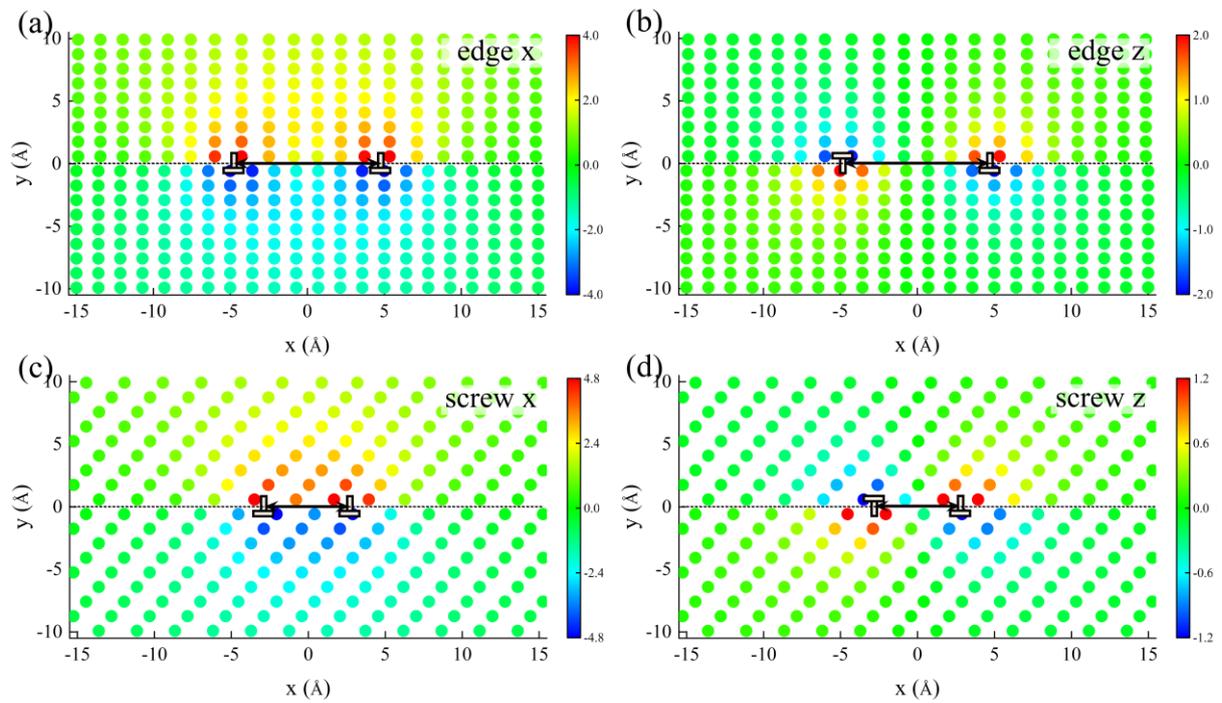

Fig. 6. Illustration of a full edge (left) or screw (right) dislocation dissociating into two Shockley partials on the (111) plane of FCC structures or (0001) plane of HCP structures. The different view directions of the pressure field in Fig. 5 for edge and screw dislocations are denoted by the red hollow arrows.

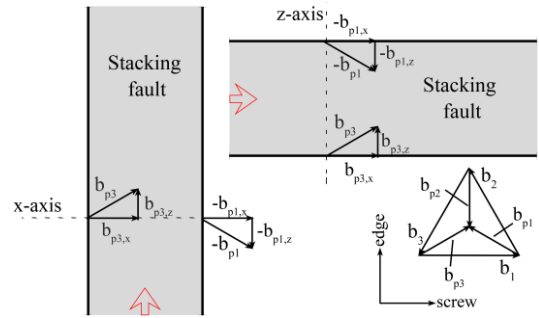

Fig. 7. The energy barrier ($\Delta E_{barrier}/c^{1/3}$, in eV) and corresponding yield stress ($\tau_{y0}/c^{2/3}$, in MPa) at 0K of edge and screw dislocations as a function of the misfit parameters $\varepsilon_b$ and $\varepsilon_s$.

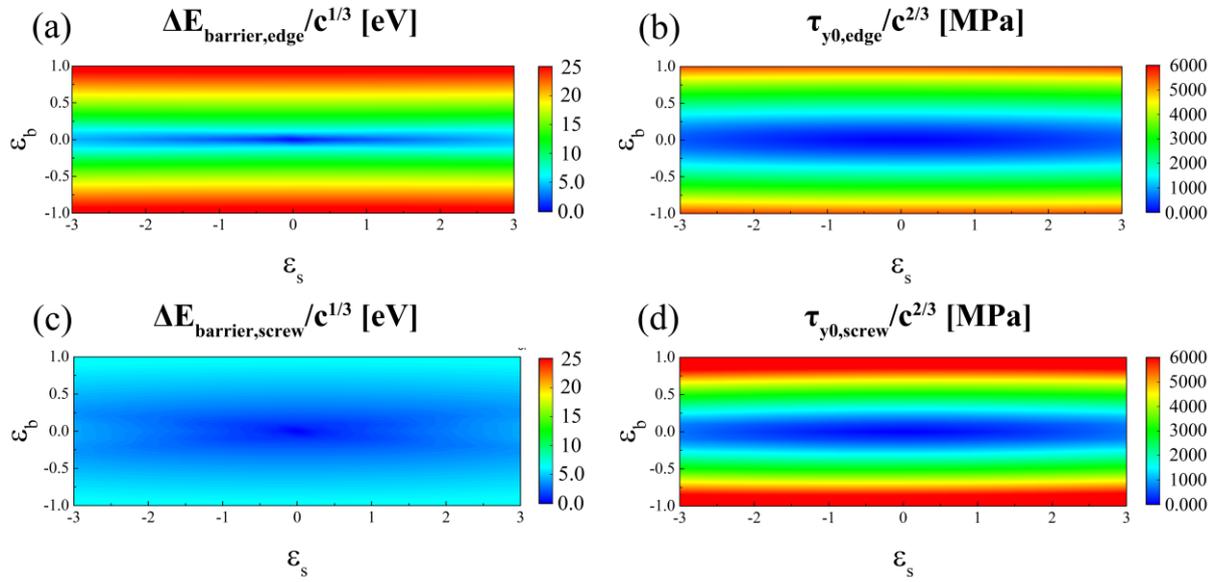

Fig. 8. (a) The correlation function $\chi(w,y_i)$ as a function of $w$ and $y_i$ for an edge dislocation in the Al-Mg alloy. (b) Normalized energy change per unit length $\Delta E_{tot}/Lc^{2/3}$ vs. the bow-out distance $w$ for an edge dislocation in the Al-Mg alloy.

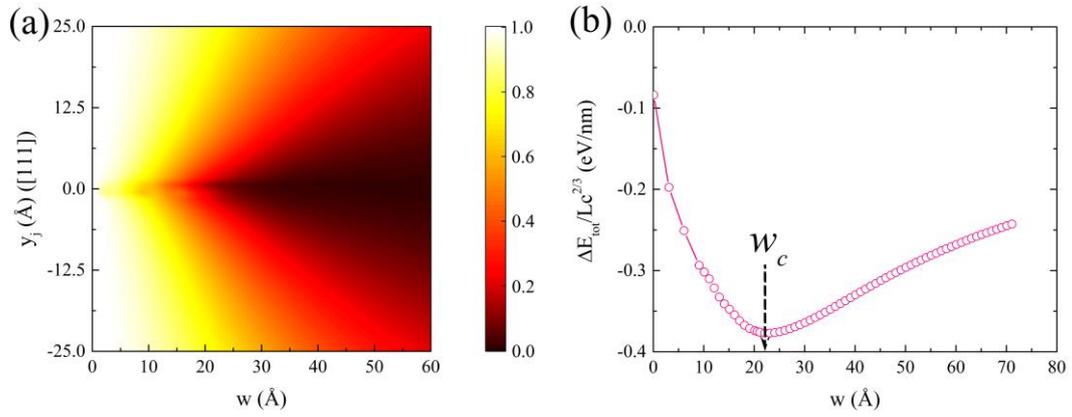